\documentclass[a4paper]{article}
\usepackage{xcolor, colortbl}
\usepackage{mathtools}
\usepackage{INTERSPEECH2022}
\usepackage{subfig}
\usepackage{placeins}
\usepackage{multirow}
\usepackage{algorithm}
\usepackage{algpseudocode}
\usepackage{tikz}
\usepackage[T1, OT1]{fontenc}

\title{Generative Data Augmentation Guided by Triplet Loss \\ for Speech Emotion Recognition}
\name{Shijun Wang$^1$, Hamed Hemati$^1$, Jón Guðnason$^2$, Damian Borth$^1$}
\address{
  $^1$University of St. Gallen, Switzerland\\
  $^2$Reykjavik University, Iceland}
\email{shijun.wang@unisg.ch, hamed.hemati@unisg.ch, jg@ru.is, damian.borth@unisg.ch}

\begin{document}

\definecolor{mygray}{RGB}{220,220,220}
\maketitle
\begin{abstract}

% Speech Emotion Recognition (SER) remains a practical problem because understanding human emotion is crucial for human-computer interaction. Two major obstacles in SER are data scarcity and data imbalance. Data scarcity is a common issue in low-resource languages, where very little labeled data is available. In contrast, data imbalance can happen in any dataset where the number of samples in different classes varies significantly. To address these issues, we exploit an augmentation strategy to improve the SER performance given imbalanced and insufficient training data. We conduct experiments and demonstrate: 1) with a highly imbalanced dataset, our augmentation strategy significantly improves the SER performance (+8\% recall score compared with the baseline); 2) in a cross-lingual benchmark, where we train a model with enough samples in the source language but very few samples in the target language, our augmentation strategy benefits the SER performance in all three target languages.

Speech Emotion Recognition (SER) is crucial for human-computer interaction but still remains a challenging problem because of two major obstacles: data scarcity and imbalance. Many datasets for SER are substantially imbalanced, where data utterances of one class (most often Neutral) are much more frequent than those of other classes. Furthermore, only a few data resources are available for many existing spoken languages. To address these problems, we exploit a GAN-based augmentation model guided by a triplet network, to improve SER performance given imbalanced and insufficient training data. We conduct experiments and demonstrate: 1) With a highly imbalanced dataset, our augmentation strategy significantly improves the SER performance (+8\% recall score compared with the baseline). 2) Moreover, in a cross-lingual benchmark, where we train a model with enough source language utterances but very few target language utterances (around 50 in our experiments), our augmentation strategy brings benefits for the SER performance of all three target languages.
\end{abstract}
\noindent\textbf{Index Terms}: speech emotion recognition, speech augmentation, cross lingual

\section{Introduction}

Speech Emotion Recognition (SER) is a task aiming to understand the underlying emotional information in speech. In recent years, there has been a surge of interest in SER, because emotion conveys crucial information during human-computer interaction. However, two obstacles usually hinder the application of SER: the imbalance and scarcity issues of the SER dataset. We can further describe the issues from two perspectives. 

Firstly, SER datasets have a common issue that the speech data distributions are non-uniform or highly skewed among different emotion classes. Speech utterances labeled as "Neutral" are much more frequent than those labeled as "Happy, Angry, etc.". The data scarcity of various emotion classes results in highly imbalanced datasets. To alleviate the data imbalance issue, a common way is to generate synthetic data through data augmentation techniques. Authors in \cite{Sahu2018OnES} use vanilla Generative Adversarial Networks (GANs) \cite{Goodfellow2014GenerativeAN} to generate synthetic openSMILE acoustic features \cite{Eyben2013RecentDI} for the training of the SER task. In \cite{Bao2019CycleGANBasedES}, the authors employ a CycleGAN \cite{Zhu2017UnpairedIT} to transfer openSMILE feature vectors from one emotion class to another. Instead of openSMILE features, some approaches do augmentation on low-level acoustic features (e.g. Mel-Spectrogram), which is an advantage for other tasks. Approaches in \cite{Rizos2020StarganFE, He2021AnIS} apply a starGAN \cite{Choi2018StarGANUG} to convert speech utterances of Neutral emotion to other emotions. Authors in \cite{Chatziagapi2019DataAU} modify a GAN to perform augmentation for imbalanced data and they empirically demonstrate that this GAN can improve the SER performance under imbalanced datasets. 

Another important aspect to consider the scarcity issue is the availability of data for various existing spoken languages. There are very few data resources available for many languages, which is a significant barrier to the research and application of SER for these languages. One way to address this problem is by training an SER model on one or more languages with enough data (source), combined with very few training utterances of other languages (target). The idea is to use the knowledge gained from source languages to improve the performance on low-resource target languages \cite{Pan2010ASO}. Such SER training strategy that spans different languages, i.e. cross-lingual SER, has been widely studied \cite{Sagha2016CrossLS, Latif2018TransferLF, Goel2020CrossLC}. Although adding source languages improves the SER performance on target languages, there is still room to improve the cross-lingual SER task. 

In this paper, We propose an augmentation strategy, which can be used to address the data imbalance and scarcity problem. One component of our augmentation approach is a GAN, whose generator is employed for augmentation. The generator is inspired by \cite{Tamkin2021ViewmakerNL}, where authors successfully perform automatic augmentation for multiple downstream tasks. To stabilize the GAN or improve its performance, additional auxiliary objectives are usually introduced during training \cite{Chen2019SelfSupervisedGV, Yamamoto2020ParallelWA, Jeong2021TrainingGW}. Thus, we apply a representation learner based on triplet loss to learn emotion representations from the original data. And the learned knowledge of the representation learner is used to increase and stabilize the augmentation performance for the SER task.

We summarize our contributions as follows: 1) We propose a GAN-based augmentation for the SER task. 2) We demonstrate that our augmentation approach can largely improve the SER performance with a highly imbalanced dataset (+8\% recall with our augmentation). 3) Moreover, for the cross-lingual SER task, our augmentation improves the performance for three different target languages with only around 50 utterances.

\section{Method}

The proposed SER data augmentation method is designed with two augmentation requirements: preserving the emotion class and providing rich variance. This means that an augmented utterance should have the same emotion label as the original one while also differing sufficiently from it. Our system setup and corresponding loss functions reflect these requirements.

% In this section, we introduce our method whose goal is to perform augmentation in accordance with two requirements. Firstly, preserving the emotion class, i.e., the original and the augmented version should belong to the same emotion class. Secondly, presenting rich augmentation variance. The augmented versions should be different even when they are from the same original version.

\subsection{Model Components} \label{sec_Model_Description}

\begin{figure}[t]
  \centering
  \includegraphics[width=0.9\linewidth]{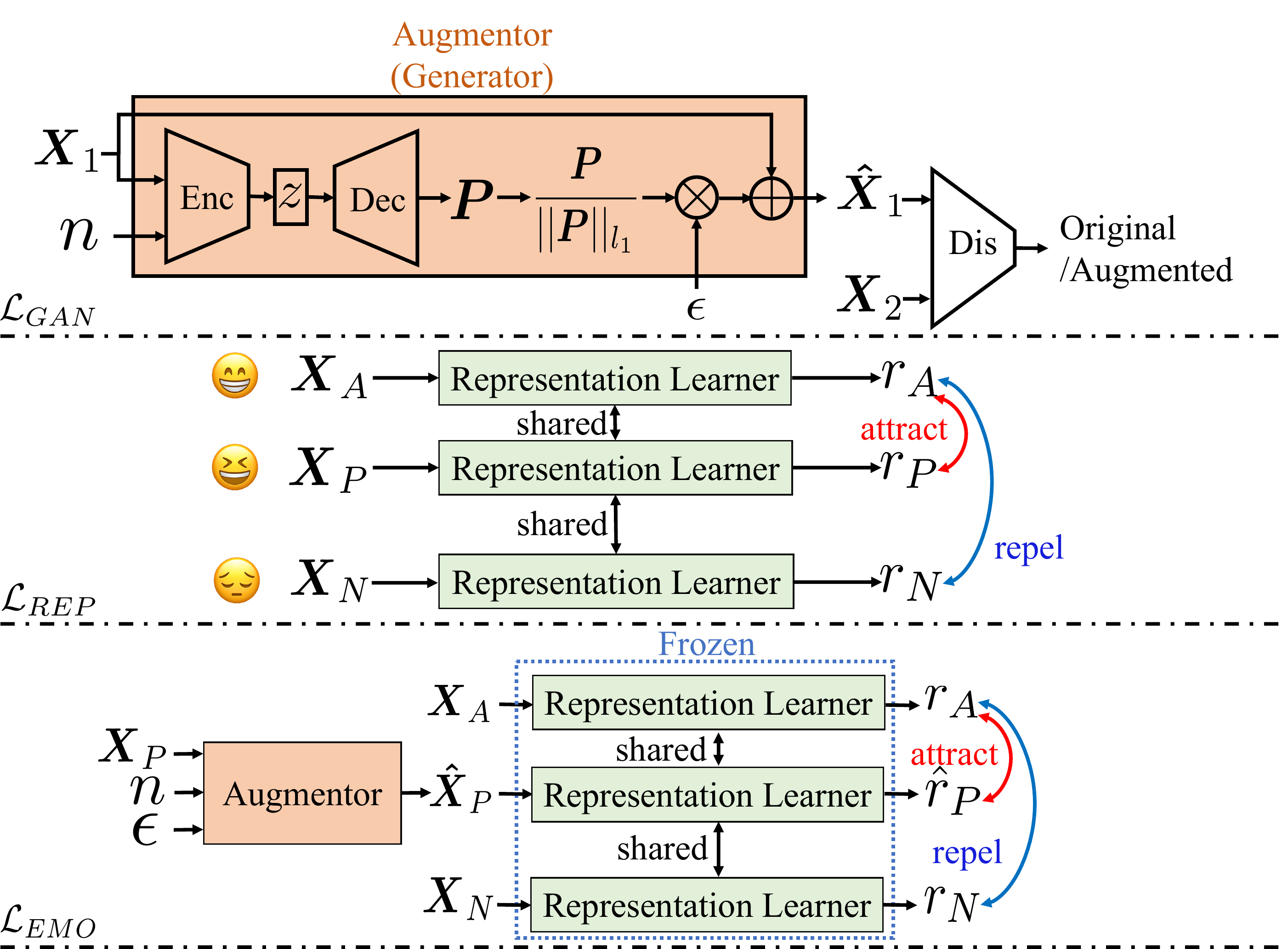}
  \caption{\textbf{TOP}: The training pipeline of our GAN. \textbf{MIDDLE}: The training pipeline of our representation learner. \textbf{BOTTOM}: The loss to guide the augmentor to preserve the emotion information.}
  \label{architecture}
  \vspace{-0.3cm}
\end{figure}

Our approach consists of a GAN and a representation learner. The generator in our GAN augment the Mel-Spectrograms. The representation learner enables the GAN to present high augmentation variance and stabilize the augmentation.

% 
% The representation learner plays a role that helps the GAN to augment the Mel-Spectrograms in a way that enables rich variance and prevents arbitrary augmentation at the same time.
% 

The GAN is shown on the top of Fig. \ref{architecture}, it contains one generator and one discriminator, while we denote the inputted Mel-Spectrograms 
as $\boldsymbol{X}$. In the rest of the paper, we refer to the generator as the augmentor since it is used to perform augmentation on Mel-Spectrograms. 
The augmentor employs an encoder to map a Mel-Spectrogram $\boldsymbol{X}_1$ into a latent vector $z$, given a Gaussian noise $\boldsymbol{n}$. Then, $z$ 
is fed to the decoder to generate a tensor $P$ with the same shape as $\boldsymbol{X}_1$. To avoid excessive removal of features from the original Mel-Spectrogram, 
we constrain the decoder output $P$ with an $l_1$-norm operation as used in \cite{Tamkin2021ViewmakerNL}. We then use a hyper-parameter $\epsilon$ to 
control the augmentation intensity and obtain the augmented Mel-Spectrogram $\boldsymbol{\hat{X}}_1$. For brevity, in this paper, we denote our
augmentor as $AUG(\boldsymbol{X})$. Finally, we feed the discriminator $D(\boldsymbol{X})$ with $\boldsymbol{X}=\boldsymbol{\hat{X}}_1$ or $\boldsymbol{X}_2$, where $\boldsymbol{X}_2$ is another 
sample from the original dataset. $D(\boldsymbol{X})$ aims to determine whether the input is augmented or not. Overall, we apply the non-saturating loss \cite{Goodfellow2014GenerativeAN} to train our GAN,
% 
% another original and the augmented Mel-Spectrograms, $\boldsymbol{X}_2$ and $\boldsymbol{\hat{X}}_1$, are fed to a discriminator $D(\boldsymbol{X})$ to determine whether they are from the original dataset or not. Overall, we apply the non-saturating loss \cite{Goodfellow2014GenerativeAN} to train our GAN,
%
\begin{equation}
  \mathcal{L}_{GAN} = \underset{AUG}{\text{min}} \underset{D}{\text{max}} 
  [ 
  \text{log} D(\boldsymbol{X}) + 
  \text{log}( -D(AUG( \boldsymbol{X})) ) 
  ].
  \label{loss_GAN}
\end{equation}
To achieve better augmentation, we further implement a representation learner. Its target is to learn meaningful emotion representations that can later be used to improve our augmentor. In contrast to a classification loss, a triplet loss is applied to train more discriminative emotion representations \cite{Harvill2019RetrievingSS, Huang2018SpeechER}. As illustrated in the middle of Fig. \ref{architecture}, our representation learner maps the Mel-Spectrogram $\boldsymbol{X}$ into an emotion representation vector $r$.
The objective is to attract the representations of $\boldsymbol{X}_{A}$ (Anchor) and $\boldsymbol{X}_{P}$ (Positive), whose emotion classes are the same, while repelling representations of $\boldsymbol{X}_{A}$ and $\boldsymbol{X}_{N}$ (Negative), since they are from different emotion classes. We define the triplet loss of our representation learner as,
\begin{equation}
    \mathcal{L}_{REP} = \text{max}(\text{dist}(r_{A},r_{P}) - \text{dist}(r_{A},r_{N})+\beta, 0),
\label{loss_REP}
\end{equation}
where dist($\cdot,\cdot$) is $l_1$ distance, and $\beta$ is the margin of the triplet loss. The margin is a hyper-parameter that defines how far away the dissimilarities should be.

The representation learner is then employed to guide the augmentor to preserve the emotion class after the augmentation. Thus, we introduce the loss $\mathcal{L}_{EMO}$ to penalize the augmentor for generating an output that is of different emotion class to the input. The setup is shown at the bottom of Fig. \ref{architecture}. As we can see, we freeze the representation learner, and an augmented Mel-Spectrogram $\boldsymbol{\hat{X}}_{P}$ is used to supersede $\boldsymbol{X}_{P}$ as one of the inputs for the triplet loss. The motivation is to use the knowledge of the representation learner to guide the augmentor to preserve emotional information. We define the loss as,
\begin{equation}
  \begin{multlined}
    \mathcal{L}_{EMO} = 
    \text{max}(\text{dist}(r_{A}, \hat{r}_{P}) - \text{dist}(r_{A},r_{N})+\beta, 0).
  \end{multlined}
\label{loss_EMO}
\end{equation}
Then, we define the model loss as,
\begin{equation}
  \begin{multlined}
    \mathcal{L}_{Model} = 
    w_{g}\mathcal{L}_{GAN} + w_{r}\mathcal{L}_{REP} + w_{e}\mathcal{L}_{EMO},
  \end{multlined}
\label{loss_M}
\end{equation}
where $\mathcal{L}_{Model}$ are weighted by $w_{g}$, $w_{r}$ and $w_{e}$.
% Now we explain how we use the representation learner to improve the augmentor for the SER task. We first need to make sure that the augmented version still belongs to the emotion class of the inputted original version. 
%
\subsection{Auxiliary Objectives for Augmentation} \label{sec_Auxiliary}
\begin{figure}[t]
  \centering
  \includegraphics[width=0.8\linewidth]{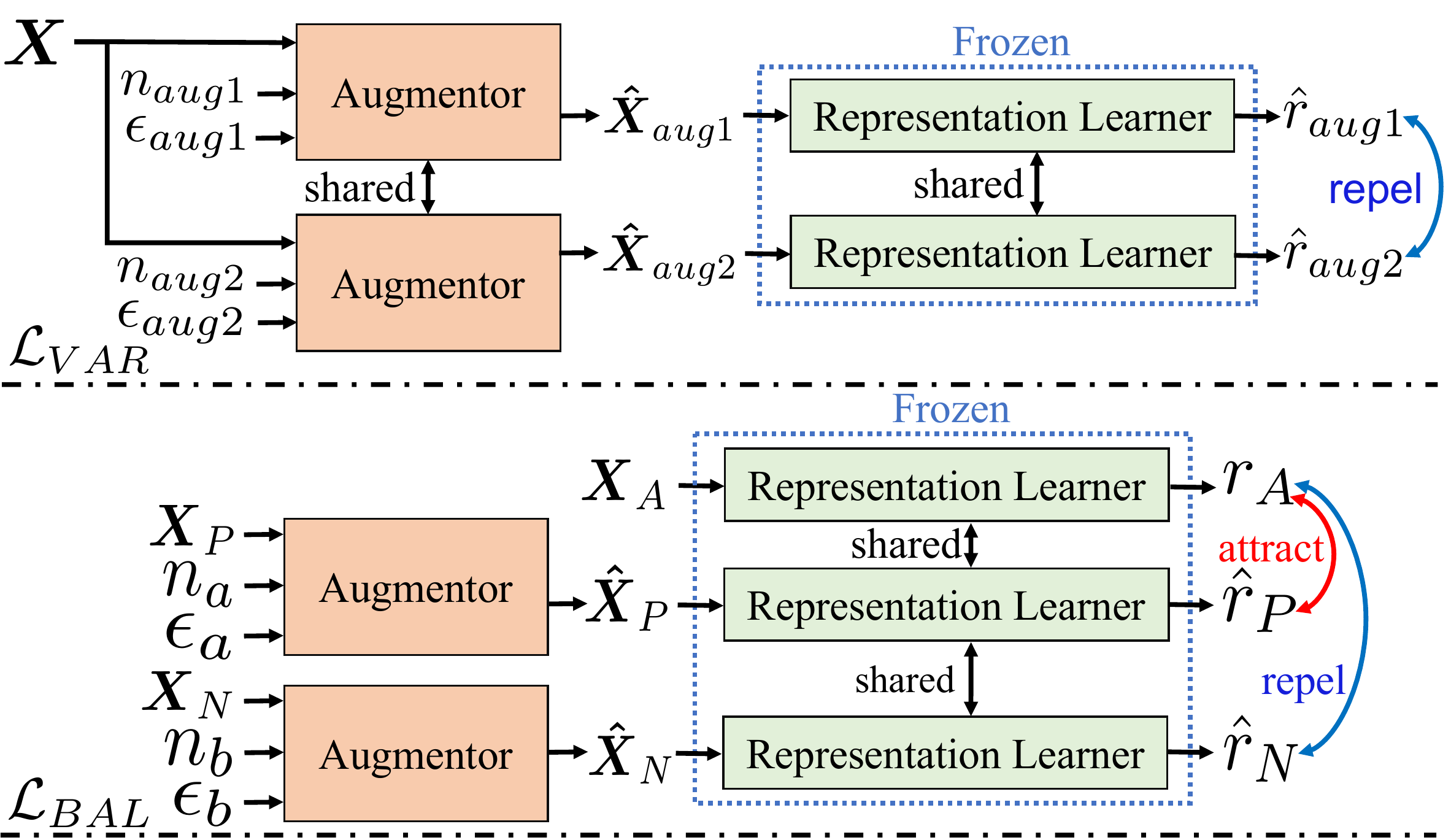}
  \caption{\textbf{TOP}: The auxiliary loss to guide the augmentor to provide augmentation variance. \textbf{BOTTOM}: The auxiliary loss to balance emotion preservation and augmentation variance.}
  \label{losses}
  \vspace{-0.4cm}
\end{figure}
To achieve better augmentation, we apply the representation learner to provide auxiliary objectives during training. We first expect the augmentor can provide  augmentation variance. In other words, for the same Mel-Spectrogram, the augmentaor can produce various augmented versions. Otherwise, identical mapping or barely perceivable augmentation of the augmentaor might not be helpful for the SER task. Thus, as shown on the top of Fig. \ref{losses}, for the same input $\boldsymbol{X}$, we first obtain two augmented Mel-Spectrograms $\boldsymbol{\hat{X}_{aug1}}$ and $\boldsymbol{\hat{X}_{aug2}}$ based on different noise $n$ and augmentation intensity $\epsilon$. We then send these two augmented Mel-Spectrograms into the representation learner to acquire representations $r_{aug1}$ and $r_{aug2}$. We freeze the representation learner as well in order to utilize its knowledge of the original data. Our target is to repel these two representations, which can force the augmentor to generate various augmented versions whose representations are dissimilar, even from the same input. We define the auxiliary variance loss as,
\begin{equation}
  \begin{multlined}
    \mathcal{L}_{VAR} = 
    \text{dot}(\hat{r}_{aug1}, \hat{r}_{aug2}).
  \end{multlined}
\label{loss_VAR}
\end{equation}
We enforce the dissimilarity by minimizing the dot product (\textup{dot($\cdot$, $\cdot$)}) between the representations from two augmented Mel-Spectrograms.

\begin{figure}[t]
  \centering
  \includegraphics[width=1.0\linewidth]{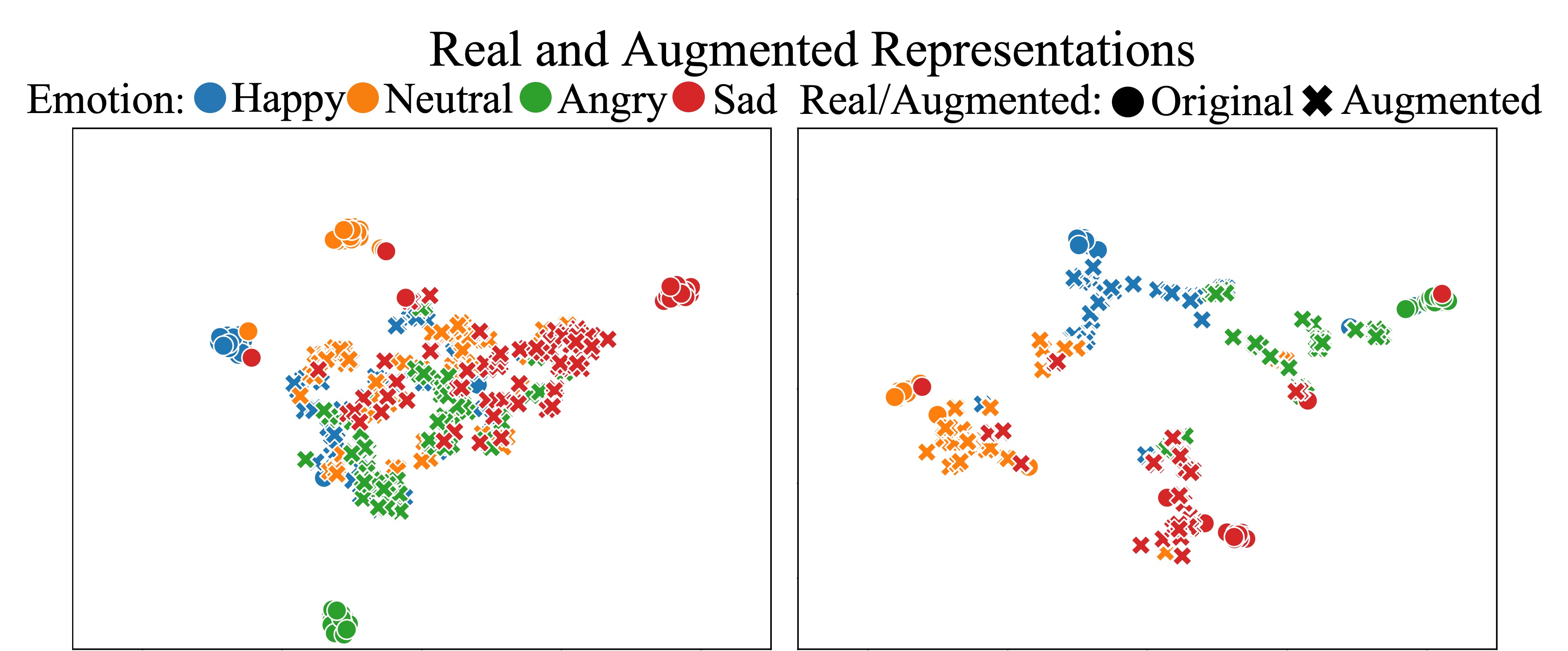}
  \caption{T-SNE visualization of the emotion representations from the representation learner. \textbf{LEFT}: the augmented utterances are from the augmentor trained without $\mathcal{L}_{BAL}$; \textbf{RIGHT}: the augmented utterances are from the augmentor trained with $\mathcal{L}_{BAL}$, which leads to a more clear cluster boundary.}
  \label{tsne}
  \vspace{-0.3cm}
\end{figure}

However, our initial experiments showed that $\mathcal{L}_{VAR}$ leads to unstable augmentation, we observe this issue by t-SNE visualisation \cite{van2008visualizing}. As shown on the left in Fig. \ref{tsne}, each point depicts the emotion representation $r$ derives from Mel-Spectrogram in the dataset IEMOCAP \cite{busso2008iemocap}. For each emotion class, we randomly pick 30 speech utterances and augment each sample 3 times with different noise and augmentation intensities. In the figure, the augmented representations from different emotions are not distinct and do not form clusters. The reason is that to meet the augmentation variance loss $\mathcal{L}_{VAR}$, the augmentor might perform arbitrarily to generate augmented versions whose representations are excessively far from original data, which means the emotion class information is damaged. Such behavior is against the loss $\mathcal{L}_{EMO}$ (Eq. \ref{loss_EMO}, aiming to preserve the emotion information), however, since the dataset is small, the augmentor is less penalized by the original data.

To enable our augmentor to preserve the emotion information while providing variant augmentations, we need to balance $\mathcal{L}_{EMO}$ (Eq. \ref{loss_EMO}) and $\mathcal{L}_{VAR}$ (Eq. \ref{loss_VAR}). Rather than doing tedious fine-tuning of the weights of these two losses, we achieve the balance by introducing another auxiliary loss as shown at the bottom of Fig. \ref{losses}. As we can see, another triplet loss is applied, but the positive and the negative inputs are both augmented. The motivation is to prevent the representations of the augmented Mel-Spectrograms from tangling together, by constraining the augmentor not only with original data, but also with augmented versions themselves. In other words, the augmented versions should be close to original versions if they belong to the same emotion, while be far from other augmented version when they from different emotion classes. Therefore, the triplet loss is defined as,
\begin{equation}
\mathcal{L}_{BAL} = \text{max}(\text{dist}(r_{A}, \hat{r}_{P}) - \text{dist}(r_{A},\hat{r}_{N})+\beta, 0).
\label{loss_BAL}
\end{equation}
As we can observe from the t-SNE visualisation on the right of Fig. \ref{tsne}, with the addition of the balancing loss, the augmented representations have a clear boundary between different classes. We conduct ablation study to demonstrate the contributions of our proposed losses further. Finally, we define the total loss as,
\begin{equation}
\mathcal{L}_{Total} = \mathcal{L}_{Model} + w_{v}\mathcal{L}_{VAR} + w_{b}\mathcal{L}_{BAL},
\label{loss_Total}
\end{equation}
where $w_{v}$ and $w_{b}$ are loss weights.
% Finally, the total loss function is a weighted sum of the introduced loss terms and these weights are defined as $w_{GAN}$, $w_{REP}$, $w_{EMO}$, $w_{VAR}$ and $w_{BAL}$. 
%
% Finally, we set the hyper-parameter loss weights $w_{GAN}$, $w_{REP}$, $w_{EMO}$, $w_{VAR}$ and $w_{BAL}$ for $\mathcal{L}_{GAN}$, $\mathcal{L}_{REP}$, $\mathcal{L}_{EMO}$, $\mathcal{L}_{VAR}$, $\mathcal{L}_{BAL}$, respectively.
%
\subsection{Training Phases}

We repeat the following steps until our models converge:

\noindent 
\textbf{1) Representation Learner:} We update our representation learner by $\mathcal{L}_{REP}$ (Eq. \ref{loss_REP}) from original data $\boldsymbol{X}$.

\noindent 
\textbf{2) Discriminator:} We freeze the augmentor, and obtain augmented Mel-Spectrograms $\boldsymbol{\hat{X}}$ from original data $\boldsymbol{X}$. The discriminator $D$ is updated by $\boldsymbol{X}$ and $\boldsymbol{\hat{X}}$.

\noindent 
\textbf{3) Augmentation Variance:} We freeze the representation learner, and utilize it to enhance the augmentation variance of the augmentor, by the loss $\mathcal{L}_{VAR}$ (Eq. \ref{loss_VAR}).

\noindent 
\textbf{4) Emotion Preservation and Balance:} We update the augmentor. We freeze the representation learner, the discriminator $D$, then generate augmented Mel-Spectrograms $\boldsymbol{\hat{X}}$ from the augmentor. The augmentor is finally updated from the discriminator loss, $\mathcal{L}_{EMO}$ (Eq. \ref{loss_EMO}) and $\mathcal{L}_{BAL}$ (Eq. \ref{loss_BAL}).

\subsection{Implementation Details}

In our augmentor, the encoder consists of 3 convolutional layers with LeakyReLU activation function and maps the input $\boldsymbol{X}$ into a latent 128D vector $z$. The decoder employs two deconvolutional layers with ReLU activation function to convert $z$ into a tensor $\boldsymbol{P}$ whose shape is the same as the input. The discriminator first uses four convolutional layers with the LeakyReLU activation function to process the input, then uses an LSTM and an attention layer \cite{Zhang2019AttentionaugmentedEM} to discern whether the input is original or augmented. Our representation learner employs 5 convolutional layers with LeakyReLU and an LSTM to process the Mel-Spectrogram, then the same attention layer in the discriminator is used to produce the 128 dimensional emotion representation.

For augmentation intensity $\epsilon$, we randomly sample $\epsilon$ from a uniform distribution $\text{U}(0.05, 0.3)$, because we found a random $\epsilon$ performs better than a fixed one as in \cite{Tamkin2021ViewmakerNL}. We set the $\beta$ in $\mathcal{L}_{REP}$ (Eq. \ref{loss_REP}), $\mathcal{L}_{EMO}$ (Eq. \ref{loss_EMO}) and $\mathcal{L}_{BAL}$ (Eq. \ref{loss_BAL}) as 7. For loss weights, we set $w_{g}$, $w_{r}$ and $w_{v}$ to 1, $w_{e}$ to 10 and $w_{b}$ to 8. We train all of our models with Adam optimizer for 30k iterations, and set the learning rate to 1e-6.

% \renewcommand{\algorithmicrequire}{\textbf{Input:}}
% \renewcommand{\algorithmicensure}{\textbf{Output:}}
% \begin{algorithm}
% \caption{Training Procedure}\label{alg}

% \begin{algorithmic}

% \Require GAN (augmentor $AUG$ and discriminator $D$), representation learner $REP$, noise $n$ drawn from a normal distribution, intensity $\epsilon$ drawn from a uniform distribution $\text{U}(0.05, 0.3)$ 

% \While{models have not converged}
% \For{each mini-batch $\boldsymbol{X}$}
%     \State Update $REP$ with $\mathcal{L}_{REP}$
%     \State Freeze $AUG$, obtain $\boldsymbol{\hat{X}}=AUG(\boldsymbol{X})$ with $n$ and $\epsilon$, update $D$ with $\boldsymbol{X}$ and $\boldsymbol{\hat{X}}$
%     \State Freeze $REP$, update $AUG$ with $\mathcal{L}_{VAR}$
%     \State Freeze $REP$ and $D$, obtain $\boldsymbol{\hat{X}}=AUG(\boldsymbol{X})$ with $n$ and $\epsilon$, update $AUG$ with discriminator loss, $\mathcal{L}_{EMO}$ and $\mathcal{L}_{BAL}$ together.
% \EndFor
% \EndWhile
% \end{algorithmic}
% \end{algorithm}

\section{Experiment and Results}

The proposed augmentation method is demonstrated with SER tasks. In short, the results show that: 1) Our augmentor can improve the SER with a highly imbalanced dataset. 2) Our augmentor can improve the cross-lingual SER task with very few training utterances from the target language. 3) the importance of the contribution of $\mathcal{L}_{VAR}$ (Eq. \ref{loss_VAR}, for adding augmentation variance), and $\mathcal{L}_{BAL}$ (Eq. \ref{loss_BAL}, for balancing the emotion preservation and providing variance). 

\subsection{Experimental Setup}

\subsubsection{Data Preprocessing and SER Classifier}

In all our experiments, we use the same acoustic features and same emotion classifier as in \cite{Chatziagapi2019DataAU}. To extract Mel-Spectrograms from waveform files, we use a 50-millisecond window and a 50 percent overlap ratio, with 128 Mel Coefficients. A VGG19 architecture \cite{Simonyan2015VeryDC} is employed as our emotion classifier, which takes a fix-sized 128 $\times$ 128 Mel-Spectrogram (128 frames) as the input and predicts the emotion class. During the evaluation, the predicted class is determined by the majority voting of segments for each Mel-Spectrogram. To train our GAN and representation learner, we use 512 $\times$ 128 Mel-Spectrograms.

\subsubsection{Datasets}

\noindent 
\textbf{Imbalanced SER} and \textbf{Ablation Study:} For the experiments of the Imbalanced SER (Sec. \ref{sec_imbalanced}) and Ablation Study (Sec. \ref{sec_ablation}) We train and test our models on IEMOCAP \cite{Busso2008IEMOCAPIE}. Like many other SER works \cite{Sahu2018OnES, Chatziagapi2019DataAU,  Latif2020AugmentingGA}, we only focus on 4 classes (Angry, Sad, Neutral and Happy), resulting in 5531 speech utterances of about 7 hours total duration. Furthermore, in order to simulate the data imbalance issue, we follow \cite{Chatziagapi2019DataAU} and randomly remove 80\% of each class except Neutral. Since IEMOCAP is originally split into 5 sessions, we conduct 5 fold cross-validation,  by using 4 sessions for training and 1 for testing. We train our SER classifier with original and augmented utterances. We augment each original training sample four times, resulting in a hybrid original-augmented training dataset that is roughly the same size as the original non-reduced dataset.

\noindent 
\textbf{Cross-Lingual SER:} For the experiment of the Cross-Lingual SER (Sec. \ref{sec_cross}), our datasets are listed in Tab. \ref{Cross-lingual Datasdet}. These datasets span four languages, with IEMOCAP and ESD serving as source languages and the remaining datasets serving as target languages. Although ESD originally has both Chinese and English speech utterances, we only use the Chinese portion in this experiment. To create IEMOCAP\_SUB and ESD\_SUB, we randomly selected utterances from IEMOCAP and ESD, in order to make all of the target language datasets the same size. Since each dataset has different class numbers, one of the consistent ways to investigate cross-lingual SER task is by considering the binary positive/negative valence classification problem, like in many other related works \cite{Sagha2016CrossLS, Latif2018TransferLF}. We follow the same method to map emotions into valance, and split 25\% of the target language datasets for our evaluation. During the training of the SER classifier, we augment each original training sample from the target language 20 times, since the target language's data is severely lacking (only 50 utterances in some experiment cases).
\vspace{-0.3cm}
\begin{figure}[t]
  \centering
  \includegraphics[width=0.9\linewidth]{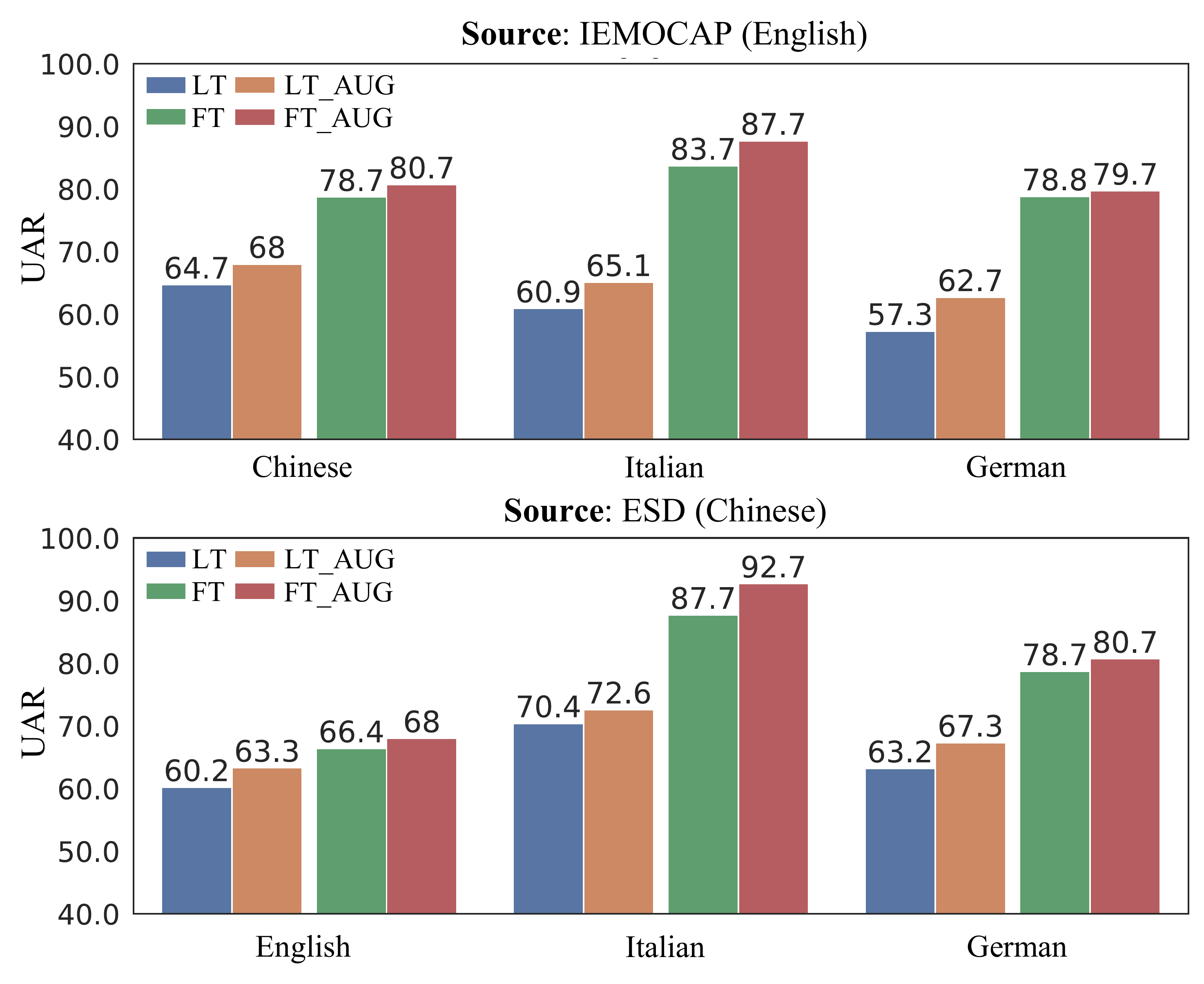}
  \caption{UAR results of cross-lingual SER. \textbf{TOP}: English (source) to 3 target languages. \textbf{BOTTOM}: Chinese (source) to 3 target languages. }
  \label{cross_lingual}
    \vspace{-0.3cm}
\end{figure}
\subsection{Results}
\subsubsection{SER on Imbalanced Dataset} \label{sec_imbalanced}
\noindent 
\textbf{Models}: We follow \cite{Chatziagapi2019DataAU} and train our SER classifier with: \texttt{NoAUG}: the 80\% reduced dataset; \texttt{AUG}: the 80\% reduced dataset and the augmented utterances.

% Despite the slight difference among the sizes of each session, we list the approximate statistic in Tab. \ref{Imbalanced_Datasdet} to make a clear description of our train and test dataset.
% 
% 
\noindent 
\textbf{Results}: We list the Unweighted Average Recall (UAR) results reported by \cite{Chatziagapi2019DataAU} and ours in Tab. \ref{Imbalanced_Result}. One thing to note is that, although using the identical setup in \cite{Chatziagapi2019DataAU} to train our \texttt{NoAUG} model, the results (the 2nd and 4th rows) differ because the 80\% removed component is randomly chosen. The average UAR results, on the other hand, are close, implying that our comparison is fair. By comparing the two \texttt{AUG} models, we can see that our augmentation strategy is roughly 5\% higher than the augmentation approach in \cite{Chatziagapi2019DataAU}, which means a better SER performance can be achieved with our augmentation approach. 
\subsubsection{Cross-Lingual SER} \label{sec_cross}
\noindent 
\textbf{Models}: We train our SER classifier with: Low Target (\texttt{LT}), 100\% of the source language data but only 10\% of the target language dataset; \texttt{LT\_AUG}, same as \texttt{LT}, but the augmented data is also fed during training; Full Target (\texttt{FT}), 100\% of the source language data and 75\% of the target; \texttt{FT\_AUG}, same as \texttt{FT}, but we add the augmented data from our augmentor.

\noindent 
\textbf{Results}: The UAR results on 25\% of the target language datasets are shown in Fig. \ref{cross_lingual}. By comparing with \texttt{LT} and \texttt{LT\_AUG}, our augmentation approach increases UAR performance by around 5\%, given only 10\% of the target language data. The performance can be further improved with the augmented data, when the target language data is additionally given, based on the results between \texttt{FT} and \texttt{FT\_AUG}.
\begin{table}[t]
  \small
  \caption{Cross-lingual datasets. IEMOCAP and ESD are used as source languages, the others serve as target languages.}
  \label{Cross-lingual Datasdet}
  \centering
  \begin{tabular}{ c | c | c | c }
    \toprule

    % \cmidrule(r){1-2}
    
    \multirow{2}*{Dataset}   & \multirow{2}*{Language}       &\multicolumn{2}{c}{Valance}  \\
    \cline{3-4}
              &                & Negative   & Positive       \\
    \midrule
    IEMOCAP \cite{Busso2008IEMOCAPIE}  & English   & 2187    & 3344 \\
    ESD \cite{Zhou2021EmotionalVC}  & Chinese  & 7000  & 7000 \\
    \midrule
    EMO-DB \cite{Burkhardt2005ADO}    & German & 385 & 150 \\   
    EMOVO \cite{Costantini2014EMOVOCA}  & Italian & 336 &252 \\
    IEMOCAP\_SUB  & English & 250 & 250 \\
    ESD\_SUB & Chinese & 250 & 250 \\
    % \midrule
    % IEMOCAP \cite{Busso2008IEMOCAPIE} & English &2187 & 3344 \\

    \bottomrule
  \end{tabular}
  \vspace{-.3cm}
\end{table}
\begin{table}[t]
  \small
  \caption{UAR results on the 80\%-reduced imbalanced dataset.}
  \label{Imbalanced_Result}
  \centering
    \begin{tabular}{ c | c | c | c | c }
    \toprule
                    & Angry   & Sad    & Happy    & Average                            \\ 
    \midrule
    NoAUG \cite{Chatziagapi2019DataAU}     & 47.8    & 46.9      & 52.2      & 49.0                               \\ 

    AUG \cite{Chatziagapi2019DataAU}      & 53.5    & 52.1      &55.2       &53.6     \\

    \midrule
    NoAUG (Ours)   &54.23        &49.03   & 47.02     & 50.09        \\ 

    \textbf{AUG (Ours)}      & \textbf{62.75}  &\textbf{58.11} &\textbf{53.71} &\textbf{58.19} \\

    \bottomrule
    \end{tabular}
    \vspace{-0.2cm}
\end{table}
\begin{table}[t!]
  \small
  \caption{Ablation study results (UAR) of the contribution of $\mathcal{L}_{VAR}$ and $\mathcal{L}_{BAL}$. }
  \label{ablation}
  \centering
  \begin{tabular}{ c | c | c | c | c }
    \toprule

    % \cmidrule(r){1-2}
    
    Model   & Angry      & Sad        &Happy  & Average\\

    \midrule
    NoAUG       &54.23        &49.03   & 47.02     & 50.09 \\
    \midrule
    $\mathcal{L}_{Model}$                   & 56.03                       & 48.43                            & 50.71  &51.52          \\
    $\mathcal{L}_{Model}+\mathcal{L}_{VAR}$                   & 36.47                       & 46.61                            & 45.31  &42.80          \\
    $\mathcal{L}_{Total}$                   & \textbf{62.75}  &\textbf{58.11} &\textbf{53.71} &\textbf{58.19}          \\

    \bottomrule
  \end{tabular}
  \vspace{-0.2cm}
\end{table}
\subsubsection{Ablation Study} \label{sec_ablation}

\noindent 
\textbf{Models}: For ablation study, we use the $80\%$ reduced dataset to train an SER classifier with and without augmentation. To train the augmentor, we consider three cases: 1) with only $\mathcal{L}_{Model}$; 2) with $\mathcal{L}_{Model}$ and auxiliary variance loss $\mathcal{L}_{VAR}$; 3) with the total loss that consists of both auxiliary terms $\mathcal{L}_{VAR}$ and $\mathcal{L}_{BAL}$.

% \texttt{AUG}: the 80\% reduced dataset; $\mathcal{L}_{Model}$: the same  and the augmented data, while the  augmentor is trained without $\mathcal{L}_{VAR}$ and $\mathcal{L}_{BAL}$; [$\mathcal{L}_{Model}+\mathcal{L}_{VAR}$]: the 80\% reduced dataset and the augmented data, while the augmentor is trained with $\mathcal{L}_{VAR}$ but without $\mathcal{L}_{BAL}$; $\mathcal{L}_{Total}$: the 80\% reduced dataset and the augmented data, while the augmentor is trained with both $\mathcal{L}_{VAR}$ and $\mathcal{L}_{BAL}$.
% \textbf{Models}: We train our SER classifier with: \texttt{AUG}: the 80\% reduced dataset; $\mathcal{L}_{Model}$: the same  and the augmented data, while the  augmentor is trained without $\mathcal{L}_{VAR}$ and $\mathcal{L}_{BAL}$; [$\mathcal{L}_{Model}+\mathcal{L}_{VAR}$]: the 80\% reduced dataset and the augmented data, while the augmentor is trained with $\mathcal{L}_{VAR}$ but without $\mathcal{L}_{BAL}$; $\mathcal{L}_{Total}$: the 80\% reduced dataset and the augmented data, while the augmentor is trained with both $\mathcal{L}_{VAR}$ and $\mathcal{L}_{BAL}$.

\noindent 
\textbf{Results}: The UAR results are listed in Tab. \ref{ablation}. We can observe that without $\mathcal{L}_{VAR}$ and $\mathcal{L}_{BAL}$, performance can only improve little because the assistance of an augmentor with low augmentation variance is limited. After we apply the loss $\mathcal{L}_{VAR}$, as we mentioned in Sec. \ref{sec_Auxiliary}, The augmentor arbitrarily performs augmentation and destroys the emotion information, resulting in a significant fall in performance. Lastly, by introducing $\mathcal{L}_{BAL}$, we can obtain the best result by balancing augmentation variance and emotion preservation. 

\section{Conclusion}

This work presents a GAN-based augmentation approach to alleviate the data imbalance and scarcity issue for the SER task. Specifically, we conduct experiments and demonstrate: 1) Even with a severely imbalanced dataset, our augmentation approach can significantly increase SER performance. 2) With only about 50 training utterances of the target languages provided, our augmentation approach considerably enhances SER performance for these low-resource languages.

\bibliographystyle{IEEEtran}

\bibliography{bib}

% \begin{thebibliography}{9}
% \bibitem[1]{Davis80-COP}
%   S.\ B.\ Davis and P.\ Mermelstein,
%   ``Comparison of parametric representation for monosyllabic word recognition in continuously spoken sentences,''
%   \textit{IEEE Transactions on Acoustics, Speech and Signal Processing}, vol.~28, no.~4, pp.~357--366, 1980.
% \bibitem[2]{Rabiner89-ATO}
%   L.\ R.\ Rabiner,
%   ``A tutorial on hidden Markov models and selected applications in speech recognition,''
%   \textit{Proceedings of the IEEE}, vol.~77, no.~2, pp.~257-286, 1989.
% \bibitem[3]{Hastie09-TEO}
%   T.\ Hastie, R.\ Tibshirani, and J.\ Friedman,
%   \textit{The Elements of Statistical Learning -- Data Mining, Inference, and Prediction}.
%   New York: Springer, 2009.
% \bibitem[4]{YourName17-XXX}
%   F.\ Lastname1, F.\ Lastname2, and F.\ Lastname3,
%   ``Title of your INTERSPEECH 2022 publication,''
%   in \textit{Interspeech 2022 -- 23\textsuperscript{rd} Annual Conference of the International Speech Communication Association, September 18-22, Incheon, Korea, Proceedings, Proceedings}, 2022, pp.~100--104.
% \end{thebibliography}

\end{document}